\documentclass[traditabstract]{aa}
\usepackage[utf8]{inputenc}
\usepackage{graphicx}
\usepackage{longtable}

\usepackage{natbib}
\bibpunct{(}{)}{;}{a}{}{,} 

\usepackage{cprotect}
\usepackage{gensymb} 
\def\m2s2{\hbox{\,m$^{2}$\,s$^{-2}$}} 
\def\Msun{\hbox{$M_{\odot}$}}             
\def\Rsun{\hbox{$R_{\odot}$}}
\def\Mjup{\hbox{$\mathrm{M}_{\rm J}$}}

\def\ten[#1]{$\;\times 10^{#1}$}

\def\solar {\ifmmode_{\mathord\odot} \else $_{\mathord\odot}$ \fi}
\def\jup {\ifmmode_{\mathrm{Jup}} \else $_{\mathrm{Jup}}$ \fi}
\def\earth {\ifmmode_{\mathord\oplus} \else $_{\mathord\oplus}$ \fi}
\def\Msol {\ifmmode {\,\mathrm{M}\solar} \else \,M\solar \fi}     
\def\Rsol {\ifmmode {\,\mathrm{R}\solar} \else R\solar \fi}     
\def\Lsol {\ifmmode {\,\mathrm{L}\solar} \else L\solar \fi}     
\def\Mjup {\ifmmode {\,\mathrm{M}\jup} \else M\jup \fi}
\def\Mearth {\ifmmode {\,\mathrm{M}\earth} \else M\earth \fi}
\def\Rearth {\ifmmode {\,\mathrm{R}\earth} \else R\earth \fi}

\def\mps {\ifmmode {\,\mathrm{m\,s^{-1}}} \else $\mathrm{m\,s^{-1}}$ \fi}     

\begin{document}

\title{Radial velocity follow-up of GJ1132 with HARPS\thanks{Based on observations made with the HARPS instrument on the ESO 3.6 m telescope under the program IDs 191.C-0873(A), and 198.C-0838(A) at Cerro La Silla (Chile). Radial velocity data will be available in electronic form at the CDS via anonymous ftp to cdsarc.u-strasbg.fr (130.79.128.5) or via http://cdsweb.u-strasbg.fr/cgi-bin/qcat?J/A+A/}}
\subtitle{A precise mass for planet 'b' and the discovery of a second planet}
\author{X.~Bonfils   \inst{\ref{ipag}} 
   \and J.-M.~Almenara \inst{\ref{ipag}}
   \and R.~Cloutier \inst{\ref{can1}, \ref{can2}, \ref{can3}}
   \and A.~W\"unsche \inst{\ref{ipag}}
   \and N.~Astudillo-Defru \inst{\ref{obsge}, \ref{concep}}
   \and Z.~Berta-Thompson \inst{\ref{cfa}, \ref{kavli}}
   \and F.~Bouchy \inst{\ref{obsge}}
   \and D.~Charbonneau \inst{\ref{cfa}}
   \and X.~Delfosse \inst{\ref{ipag}}
   \and R.~F.~D\'iaz \inst{\ref{uba}, \ref{iafe}}
   \and J.~Dittmann \inst{\ref{kavli}}
   \and R.~Doyon \inst{\ref{can3}}
   \and T.~Forveille \inst{\ref{ipag}}
   \and J.~Irwin \inst{\ref{cfa}}
   \and C.~Lovis \inst{\ref{obsge}}
   \and M.~Mayor \inst{\ref{obsge}}
   \and K.~Menou \inst{\ref{can1}}
   \and F.~Murgas \inst{\ref{iac},\ref{ull}}
   \and E.~Newton  \inst{\ref{kavli}}
   \and F.~Pepe \inst{\ref{obsge}}
   \and N.~C.~Santos \inst{\ref{caup}, \ref{uporto}}
   \and S.~Udry \inst{\ref{obsge}}
}
	
\institute{Univ. Grenoble Alpes, CNRS, IPAG, 38000 Grenoble, France\label{ipag}
  \and Observatoire de Gen\`eve, Universit\'e de Gen\`eve, 51 ch. des Maillettes, 1290 Sauverny, Switzerland\label{obsge}
  \and Dept. of Astronomy \& Astrophysics, University of Toronto, 50 St. George Street, M5S 3H4, Toronto, ON, Canada\label{can1}
  \and Centre for Planetary Sciences, Dept. of Physical \& Environmental Sciences, University of Toronto Scarborough, 1265 Military Trail, M1C 1A4, Toronto, ON, Canada\label{can2}
  \and Institut de Recherche sur les Exoplan\`etes, d\'epartement de physique, Universit\'e de Montr\'eal, C.P. 6128 Succ. Centre-ville, H3C 3J7, Montr\'eal, QC, Canada\label{can3}
  \and Universidad de Concepci\'on, Departamento de Astronom\'ia, Casilla 160-C, Concepci\'on, Chile\label{concep}
  \and Harvard-Smithsonian Center for Astrophysics, 60 Garden Street, Cambridge, Massachusetts 02138, USA\label{cfa}
  \and Kavli Institute for Astrophysics and Space Research, Massachusetts Institute of Technology, 77 Massachusetts Avenue, Cambridge, Massachusetts 02139, USA\label{kavli}
  \and Universidad de Buenos Aires, Facultad de Ciencias Exactas y Naturales. Buenos Aires, Argentina.\label{uba}
  \and CONICET - Universidad de Buenos Aires. Instituto de Astronom\'ia y F\'isica del Espacio (IAFE). Buenos Aires, Argentina.\label{iafe}
  \and Instituto de Astrof\'{i}sica e Ci\^encias do Espa\c{c}o, Universidade do Porto, CAUP, Rua das Estrelas, 4150-762 Porto, Portugal\label{caup}
  \and Departamento de F\'{i}sica e Astronomia, Faculdade de Ci\^encias, Universidade do Porto, Rua do Campo Alegre, 4169-007 Porto, Portugal\label{uporto}
  \and Instituto de Astrof\'sica de Canarias (IAC), E-38200 La Laguna, Tenerife, Spain\label{iac}
  \and Dept. Astrof\'isica, Universidad de La Laguna (ULL), E-38206 La Laguna, Tenerife, Spain\label{ull}
}

\date{xxx 2017}

\abstract{GJ1132 is a nearby red dwarf known to host a transiting Earth-size planet. After its initial detection, we pursued an intense follow-up with the HARPS velocimeter. We now confirm the detection of GJ1132b with radial velocities only. We refined its orbital parameters and, in particular, its mass ($m_b = 1.66\pm0.23 M_\oplus$), density ($\rho_b = 6.3\pm1.3$ g.cm$^{-3}$) and eccentricity ($e_b < 0.22 $; 95\%). We also detect at least one more planet in the system. GJ1132c is a super-Earth with period $P_c = 8.93\pm0.01$ days and minimum mass $m_c \sin i_c = 2.64\pm0.44~M_\oplus$. Receiving about 1.9 times more flux than Earth in our solar system, its equilibrium temperature is that of a temperate planet ($T_{eq}=230-300$ K for albedos $A=0.75-0.00$) and places GJ1132c near the inner edge of the so-called habitable zone. Despite an a priori favourable orientation for the system, $Spitzer$ observations reject most transit configurations, leaving a posterior probability $<1\%$ that GJ1132c transits. GJ1132(d) is a third signal with period $P_d = 177\pm5$ days attributed to either a planet candidate with minimum mass $m_d \sin i_d = 8.4^{+1.7}_{-2.5}~M_\oplus$ or stellar activity. Its Doppler signal is the most powerful in our HARPS time-series but appears on a time-scale where either the stellar rotation or a magnetic cycle are viable alternatives to the planet hypothesis. On the one hand, the period is different than that measured for the stellar rotation ($\sim125$ days) and a Bayesian statistical analysis we performed with a Markov Chain Monte Carlo and Gaussian Processes demonstrates the signal is better described by a keplerian function than by correlated noise. On the other hand, periodograms of spectral indices sensitive to stellar activity shows power excess at similar periods than that of this third signal, and RV shifts induced by stellar activity can also match a keplerian function. Eventually, we prefer to leave the status of 'd' undecided.
}


\maketitle

\section{Introduction}
GJ1132 is an M dwarf of our solar neighborhood with a known transiting planet detected by the MEarth survey \citep{2015Natur.527..204B}. Owing to the small size, low mass and low temperature of the parent star ($\sim$ 0.21 R$_\odot$, 0.18 M$_\odot$, 3'300 K), the 1.6-day periodic, 2.6-mmag dips observed in its photometric light curves imply the planet has a size comparable to that of Earth ($\sim$ 
1.2~R$_\oplus$) and a warm equilibrium temperature ($\sim$ 400-600 K). Being 2-3 mag brighter than most other Earth-size planet hosts seen by, e.g., Kepler, GJ\,1132 is an appealing system for follow-up characterization \citep{2017ApJ...850..121M, 2017AJ....153..191S}.

The discovery paper already includes a radial-velocity time-series (RV) collected with the HARPS spectrograph. With an orbital model composed of a single planet and a fixed zero eccentricity we measured an orbital semi-amplitude of 2.76$\pm$0.92 $ \rm m\,s^{-1}$, corresponding to a planetary mass of $1.62\pm0.55$~M$_\oplus$ \citep{2015Natur.527..204B}. Although it favours a rocky composition the constraint is loose given the large mass uncertainty; to the point that even a gaseous composition remains possible in a 3-$\sigma$ range. In addition to bulk composition, the mass of the planet is also an important parameter to determine the scale height of the atmosphere and constrain transmission spectroscopy observations \citep{Schaefer:2016cc, 2017AJ....153..191S}.

If a better mass measurement already makes a strong case for pursuing an intensive radial-velocity follow-up, searching for additional planets is an equally good motivation. With a known transiting planet, we know the system is favourably aligned, and that additional planets have a high chance of being seen in transit as well \citep{2011A&A...525A..32G, 2017Natur.542..456G}. Actually, GJ1132 was recently used in simulations as an illustrative example of such a strategy \citep{2017AJ....153....9C}.

This paper reports on our radial-velocity follow-up campaign on GJ1132 with the HARPS spectrograph. We identify a first component with period $P_d=177\pm 5$ days that we attribute to either an outer planet with mass $m_d \sin i=8.4^{+1.7}_{-2.5}~M_\oplus$ or stellar activity. After subtracting this first signal, we show GJ1132b is now identified with the sole radial-velocity data. Our time series then reveals another planet with mass $m_c \sin i_c = 2.7\pm0.4~M_\oplus$ and period $P_c = 8.92\pm0.01$ days. With an equilibrium temperature of 230-300 K, that super-Earth is located near the habitable zone. Matching our ephemeris with $Spitzer$ observations from \citet{2017AJ....154..142D}, we found that, unfortunately, transits of planet $c$ are largely excluded. 
 
\section{Data}
From June 6th, 2015 (BJD=2457180.5) to June 21st, 2017 (BJD=2457925.5), we collected 128 observations with the HARPS spectrograph \citep{2003Msngr.114...20M, 2004A&A...423..385P}, including the 25 measurements already published in \citet{2015Natur.527..204B}. We chose the high-resolution mode (R=115'000), used the scientific fiber for the target and the calibration fiber for the sky. In practice, sky brightness is low enough and the second fiber is not used by our pipeline. It only serves as a potential $a~posteriori$ diagnostic. Exposure times were fixed to 40 minutes, except for one exposure on June 26th, 2016 (BJD=2457566.5) which was shortened to 1700 sec. Also, note that we discarded a 90th measurement otherwise found in ESO archives. It was just 5 sec and taken on June 11th, 2015 (BJD=2457185.5). 

The online pipeline produces extracted spectra calibrated in wavelength \citep{2007A&A...468.1115L}. It also computes radial velocities by doing the cross-correlation with a numerical mask \citep{Baranne:1996tb, 2002A&A...388..632P}. We use this initial estimate to shift all spectra to a common reference frame, and we co-add them to build a high signal-to-noise reference spectrum. We then refine the radial-velocity determination by finding the best Doppler shift between this reference template and individual spectra \citep[e.g.][]{1997MNRAS.284..265H, 2006A&A...447..355G, AngladaEscude:2012gd, 2015A&A...575A.119A}. 

The radial-velocity uncertainties are evaluated by measuring the Doppler information content in the $\chi^2(RV)$ profile, using the formalism of \citet{2001A&A...374..733B} and \citet{2010A&A...523A..88B}. That formalism quantifies the radial-velocity uncertainty by doing a weighted sum over the spectral elements with more weight to the spectral elements with a higher derivative. And since the derivative of a spectrum has a higher variability against noise than the spectrum itself, we instead apply the formula directly to the $\chi^2(RV)$ profile, which has a few hundred times higher signal-to-noise. For GJ1132, a V=13.5 mag star, we estimate that photon-noise contributes to 2-3 $\rm m\,s^{-1}$ to the precision of individual measurements.

In addition to radial velocities, we also measure several activity proxies such as spectroscopic indices (H$_\alpha$, H$_\beta$, calcium S index and Na; see \citet{2017A&A...600A..13A}). Spectroscopic indices  can reveal and trace inhomogeneities at the surface of the star which, animated by the stellar rotation, can contribute to an apparent Doppler shift unrelated to the presence of planets \citep{2007A&A...474..293B, 2014Sci...345..440R}.

Our time-series for the RVs, their uncertainties, and for activity proxies is reported in Table~\ref{tab:rv}. 

\begin{figure*}
\includegraphics[width=\linewidth]{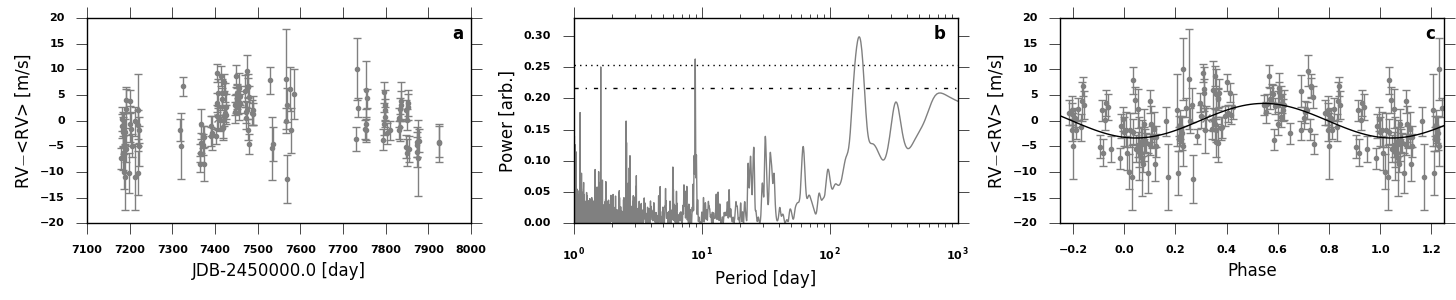}\\
\includegraphics[width=\linewidth]{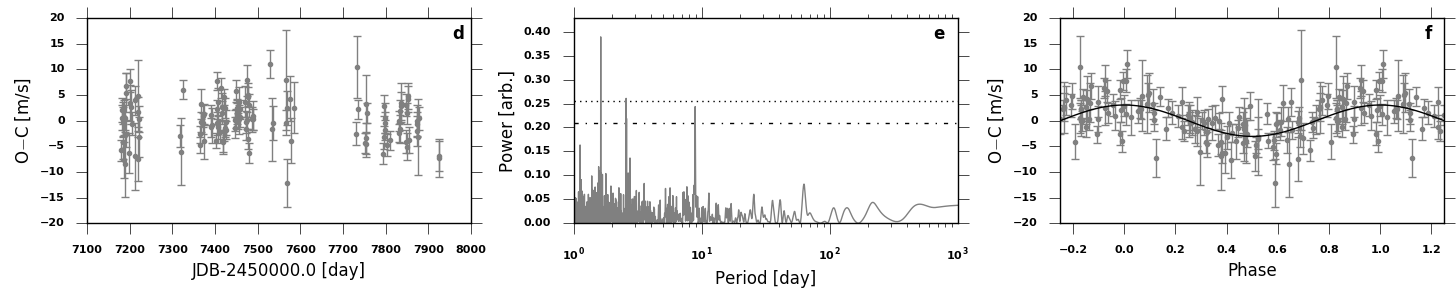}\\
\includegraphics[width=\linewidth]{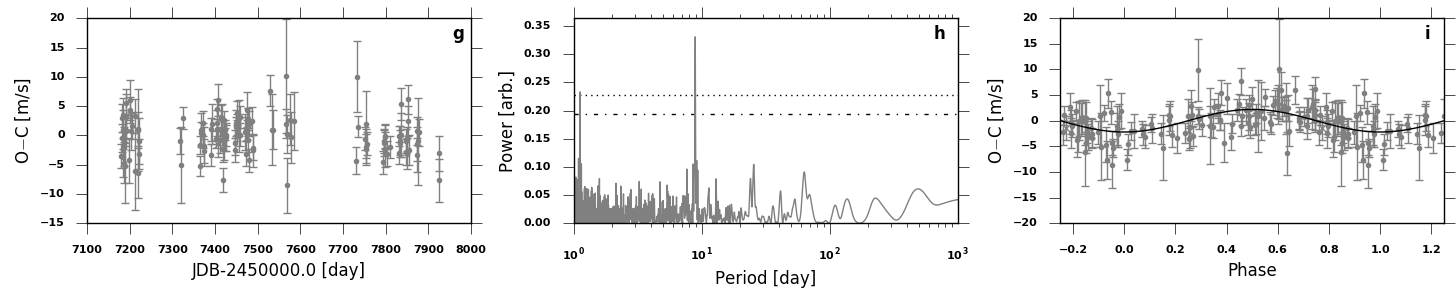}\\
\caption{\label{fig1}{\bf Iterative periodogram analysis.} {\it Left column:} RV times-series, before any subtraction {\bf (a)}, after subtracting 1 keplerian fit {\bf (d)} and after subtraction of a 2-keplerian fit {\bf (g)}. {\it Middle column :} Periodograms for each RV time-series on the left column. {\it Right column:} RV time-series seen on the left column phase folded to the period of maximum power seen in the periodograms of the middle column. The best sine-fit is superimposed. Dash-otted (resp. dotted) lines in panels $b$, $c$, $e$ and $h$ are placed at a power level corresponding to false-alarm probability of 1\% (resp. (0.1\%).
}
\end{figure*}

\section{\label{sect:analysis}Analysis}

\subsection{Iterative periodogram analysis}
We start with the RV time series (Fig.~\ref{fig1}a) and compute its Generalized Lomb-Scargle periodogram \citep[Fig.~\ref{fig1}b;~][]{Press:1992vz, 2009A&A...496..577Z}. We choose a normalization such that a power of 1 at a given period means that a sin-wave fit to the data is a perfect fit ($\chi^2=1$) whereas a power of 0 means a sin-wave fit does not improve the $\chi^2$ over that of a fit by a constant. We identify several peaks with a high power: one at the period of the known planet GJ\,1132b ($\sim$1.63 day), one near the period 8.9 day, and, the strongest, around 171 days. They have power p=0.25, 0.26 and 0.30, respectively. 

To evaluate the significance as a function of the power excess that is measured, we create synthetic data illustrative of time-series with noise only. To preserve the sampling the simulated time-series have the same dates as the original. To preserve the distribution of RVs around their mean, the values are picked by randomly shuffling the original time-series. We compute the periodogram of simulated time-series and measure their power maxima. After many trials we can build a distribution of power maxima that serves as comparison with the power values measured in the original time series. 
We found that, over 10'000 trials, no simulated time-series has power maxima equal or higher than p=0.30, meaning that the peak seen around P=177 day has a false-alarm probability (FAP) lower than 1/10'000=0.01\%; equivalent to $>3.8\sigma$ significant.

We note that \citet{2015Natur.527..204B} found a rotation period of 125 days and that a sampling rate of $\sim 1 yr$ could produce an alias at the period 190 day, indistinguishable from the 171-day pic. Nevertheless, the signal appears well sampled when phased with a period of 177 or 190 days. and the 177-day pic is therefore unlikely an alias. 

\begin{figure}
\includegraphics[width=\linewidth]{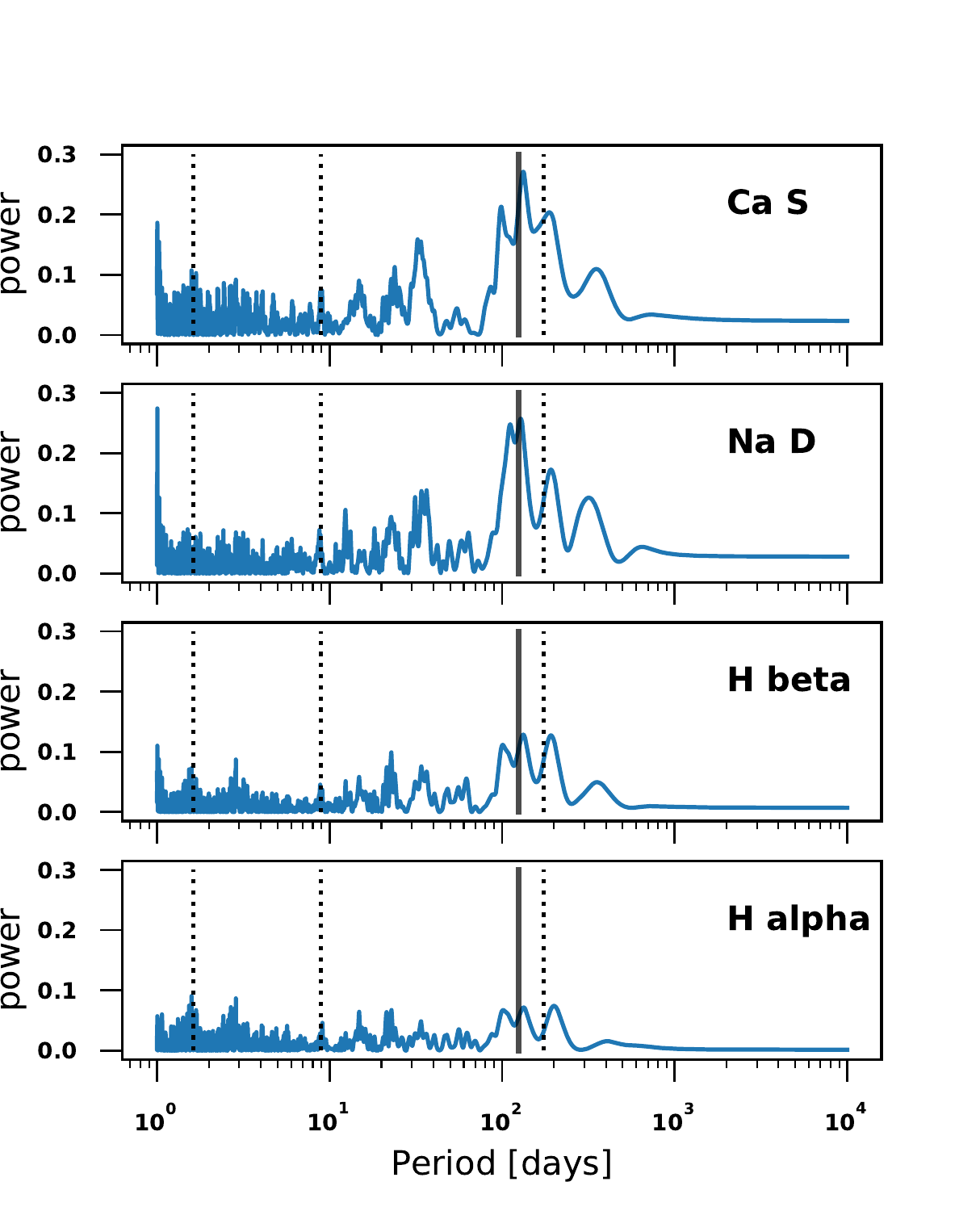}\\
\caption{\label{fig:pind} Periodogram of spectral indices measured on HARPS spectra. \ion{Ca}{}, \ion{Na}{}, \ion{H$_\beta$}{} and \ion{H$_\alpha$}{} are shown from top to bottom, respectively. The period of the stellar rotation is shown with a vertical full line and the periods of the three Doppler signals attributed to GJ1132 b, c and, d are shown with vertical dashed lines.}
\end{figure}

We next reproduce the same periodogram analysis with spectroscopic indices. Figure~\ref{fig:pind} shows these periodograms for \ion{Ca}{} (blue), \ion{Na}{} (green), \ion{H$_\beta$}{} (red) and \ion{H$_\alpha$}{} (cyan). The period of the stellar rotation is shown with the vertical full line and the periods of the three Doppler signals discussed in this paper are shown with vertical dashed lines. We see broad power excess between 80-300 days with a highest peak at the period of the stellar rotation ($\sim$125 day). Peaks of power are also seen near 175 day, calling for caution on the interpretation of the corresponding Doppler signal. If not due to stellar rotation, magnetic cycle can also induce periodic variations \citep{2011A&A...534A..30G}. In Sect.~\ref{sect:mcmc}, we perform a more detailed modelling using Markov Chain Monte Carlo and Gaussian Processes algorithms. By statistically comparing models we show the signal is best describe by a keplerian than correlated noise. Nevertheless, since stellar activity can produce RV variations that match a keplerian \citep{2007A&A...474..293B}, we do not consider this comparison definitely favors the planet interpretation.

We fit and remove a keplerian component to the RVs, and now look at the residuals (Fig.~\ref{fig1}; panel d). 
The most prominent peak is now that of GJ1132b and RV residuals are well modeled by a keplerian with period P=1.63 day (Fig.~\ref{fig1}f). The peak itself has a false-alarm probability of less than 0.01\%. The detection is now significant even without prior knowledge on the period from the photometric transits. GJ1132b is therefore confirmed from the sole radial-velocity data.

We continue with a model composed of 2 keplerian orbits. Looking at the residuals we now see strong power excess around P=8.9 days (FAP$<$0.01\%; Fig.~\ref{fig1}g), also well modeled with an additional keplerian (Fig.~\ref{fig1}i).

The rotation measured for GJ1132 is unambiguously distinct from the last two RV periodicities. This makes stellar activity an unlike culprit of these two short-period signals. We attribute them instead to two orbiting planets, namely GJ1132 b and c. 

\subsection{\label{sect:mcmc}Joint modeling of planets and correlated `noise'}
Here we apply a second, complementary analysis of the data using a
non-parametric Gaussian process (GP) regression model of the correlated RV
residuals. GP regression modeling works within a Bayesian framework providing a
distribution of functions that model the correlations between adjacent RV
measurements following the removal of a mean planetary model containing
up to 3 planets in our analysis. This technique has been used recently in the
literature to model stellar RV activity thus facilitating the detection and precise
characterization of planets around active stars
\citep[e.g.][]{Haywood:2014hs, 2015MNRAS.452.2269R, 2017AJ....153....9C, Donati:2016dr, 2018A&A...613A..25B}.
A complete description of the techniques used to simultaneously model the RV
variations with a GP plus keplerian orbital solutions can be found in
\citet{2017A&A...608A..35C}. Here we briefly summarize the key steps/assumptions
used in this work.

As mentioned, the broad peak centred around $\sim 175$ days in the LS periodogram
of the RVs spans the star's photometric rotation period of 125 days \citep{2017AJ....154..142D}. Therefore the stellar activity might be
modulated at approximately the star's rotation timescale. Thus we adopt a
quasi-periodic covariance kernel for the GP activity model of the form

\begin{equation}
k_{i,j} \propto \exp{\left[ - \frac{|t_i-t_j|^2}{2\lambda^2} -\Gamma^2
    \sin^2{\left(\frac{\pi|t_i-t_j|}{P_{\rm GP}} \right)} \right]},
\label{cov}
\end{equation}

\noindent where $t_i$ is the $i^{\rm th}$ BJD in the time-series for
$i,j=1,\dots,128$. The GP hyperparameters $a,\lambda,\Gamma,$ and $P_{\rm GP}$
describe the amplitude of the correlations, the exponential decay timescale,
the coherence scale of the correlations, and the periodic timescale
($P_{\rm GP} = P_{\rm rot}$ in photometry) respectively. 
Because the GP is intended to model only the stellar activity the
posterior probability density functions (PDFs)
of the four hyperparameters are trained via Markov Chain Monte-Carlo (MCMC) on a training
set which is independent of planetary signals. For this purpose, we used \texttt{emcee} \citep{2013ascl.soft03002F}, a \texttt{python} implementation of the affine-invariant ensemble MCMC sampler \citep{Goodman+10}.
 We opt for a training set using the MEarth photometry presented in \cite{2015Natur.527..204B} and used in
  \cite{2017AJ....153....9C} to model the stellar RV activity using a GP. Here a well-defined
  solution is easily found whose periodic term is the photometric rotation period.
The resulting marginalized posterior PDFs of the GP hyperparameters
$\lambda,\Gamma,$ and $P_{\rm GP}$ are then used as priors when next we jointly
model the RVs simultaneously with planetary signals and a quasi-periodic GP.
The posterior PDFs of these
three hyperparameters from training are each approximated by a 1-dimensional kernel
density estimation which can then be sampled during the modeling of the RVs. 
The covariance amplitude $a$ is left as an effectively
unconstrained free parameter in the RV modeling.

We then model the RVs with one of four potential planetary models. The first model contains the two planets GJ 1132 b and c. The second model contains the same two planets plus the quasi-periodic GP. The third model asummes three planets GJ 1132 b, c, and d. And the fourth model contains the same three planets plus the quasi-periodic GP. In each considered model the orbital period and time
of mid-transit of GJ 1132b are assigned Gaussian priors based on the transit
results from \cite{2015Natur.527..204B}. The orbital
period of the putative GJ 1132c (resp. GJ 1132d) is assigned a uniform prior between 8.8 and 9.0 days
days (resp.  between 120 and 220 days). We adopt uninformative Jeffreys priors
between 0-10 $m\,s^{-1}$ on each planet's semi-amplitude. This choice of prior was
modified and not found to significantly affect the results. The eccentricities
$e$ of each planet were sampled indirectly via the jump parameters 
$h=\sqrt{e} \cos{\omega}$ and $k=\sqrt{e} \sin{\omega}$ where $\omega$ is the
argument of periastron, a choice that reduces bias toward high eccentricities \citep{Ford:2006ej}.

Next, we conduct a model
comparison using time-series cross-validation \citep{2009arXiv0902.3977A}; a computationally
less expensive procedure than computing the fully marginalized Bayesian
likelihood and one that is independent of the choice of model parameter priors.
The unique model parameters for each model---including the GP
hyperparameters---
are optimized on each of the 107 training sets which contain between 20 and the full dataset size, less 1 (i.e. 127), chronologically-spaced RV measurements.
For each split of the data the
  testing set is the single measurement taken after the final measurement in the
  training set. The lnlikelihood ($\ln{\mathcal{L}}$) of the testing data given each model optimized
on the training set are then computed. Over the 107 splits of the data we compute the median and median absolute deviation per measurement of each
model's $\ln{\mathcal{L}}$. 
Table 1 reports the resulting $\ln{\mathcal{L}}$ ratio for various pairs of competing models where each model's $\ln{\mathcal{L}}$ is calculated by scaling the median $\ln{\mathcal{L}}$ per measurement from cross-validation to the full dataset size of 128 measurements.
We find that the 3 planet model is greatly favoured over the 2 planet model
with a $>8\sigma$ greater $\ln{\mathcal{L}}$. Furthermore the 3 planet $+$ GP
model has a marginally better $\ln{\mathcal{L}}$
to the 3 planet model (i.e. $0.3\sigma$ higher $\ln{\mathcal{L}}$). 
We therefore conclude that the more simplistic model with 3 keplerian signals and no GP is the model most heavily favoured by the
data. Although it favors the planet interpretation for GJ1132(d), we refrain from a definitive conclusion. Indeed, as shown by previous examples from our survey, activity-induced Doppler shifts can match a keplerian signal \citet{2007A&A...474..293B}.

\begin{table}
 \caption{ \label{tab:odds}Maximum $\ln{\mathcal{L}}$ ratios for various competing models}
\tiny
  \renewcommand{\arraystretch}{0.7}
\centering
\label{table}
\begin{tabular}{cccc}
\hline \\ [-1ex]
\multicolumn{3}{c}{Model comparisons} & lnLikelihood ratio \smallskip \\
\hline \\ [-1ex]
3 keplerians &vs.& 2 keplerians & 9e17 ($>$8$\sigma$)\\
2 keplerians $+$ GP& vs.& 2 keplerians & 7e13 (7.6$\sigma$)\\
3 keplerians &vs. & 2 keplerians $+$ GP & 1e4 (3.9$\sigma$)\\
3 keplerians $+$ GP& vs. &2 keplerians $+$ GP & 1e5 (4.4$\sigma$)\\
3 keplerians &vs.&  3 keplerians $+$ GP & 1.34 (0.3$\sigma$)\\
\hline
\end{tabular}
\end{table}

The resulting RV model parameters, assuming that the observed RV variations are due to 3 planets
plus residual stellar activity which we model with a non-parametric GP, are reported in
Table~\ref{table:results}. GJ1132 (d) is presented with parenthesis around its $d$ letter to stress it is not accepted as a planet detection. 

In the MCMC from which these results were derived we initialized 200 walkers and ran each chain
for a duration of approximately 20 autocorrelation times to ensure adequate convergence of the
chains. The steps corresponding to the first $\sim 10$ autocorrelation times were treated as
the burn-in phase and discarded. The results are broadly consistent with the results from
the iterative periodogram analysis of Sect. 3.1.
The resulting marginalized posterior PDFs of the model parameters are shown in Fig.~\ref{A1}.

As a complement, Figures~\ref{A2} and \ref{A3} show phase-folded RVs and residuals with GP regression removed.
 The uncertainties are not rescaled. After removing the mean GP model, and the 
best-fit keplerians for GJ1132 b and c, the residual rms is 2.74 m/s.

\begin{table*}
\caption{\label{table:results}Model parameters. Maximum a posteriori and 68.3\% confidence intervals. Planets $b$ and $c$ are considered robust detections, whereas planet $(d)$ is considered a planet candidate, and possibly the result of stellar activity.}
\tiny
  \renewcommand{\arraystretch}{0.7}
\centering
\label{table}
\begin{tabular}{lccc}
\hline \\ [-1ex]
Parameter & \multicolumn{3}{c}{Maximum a-posteriori values with $16^{\rm{th}}$ and $84^{\rm{th}}$ percentiles} \smallskip \\
\hline \\ [-1ex]
\emph{Stellar Parameters} & & \smallskip \\
Stellar mass, $M_s$ [$M_{\odot}$]         & \multicolumn{3}{c}{0.181 $\pm$ 0.019}  \\
Stellar radius, $R_s$ [$R_{\odot}$]       & \multicolumn{3}{c}{$0.2105^{+0.0102}_{-0.0085}$}  \\
Stellar Luminosity, $L_s$ [$L_{\odot}$]     & \multicolumn{3}{c}{0.00438 $\pm$ 0.00034} \\
Effective Temperature, $T_{\rm{eff}}$ [K] & \multicolumn{3}{c}{3270 $\pm$ 140} \\
Rotation Period, $P_{\rm{rot}}$ [days] & \multicolumn{3}{c}{$122.3^{+6.0}_{-5.0}$} \\
Systemic velocity, $\gamma_0$ [m s$^{-1}$] & \multicolumn{3}{c}{35078.8 $\pm$ 0.8} \medskip \\

\emph{Gaussian Process Hyperparameters} & & \smallskip \\
ln Correlation amplitude, $\ln{a}$ [m s$^{-1}$]    & \multicolumn{3}{c}{$-0.18^{+1.12}_{-1.32}$}  \\
ln Exponential timescale, $\ln{\lambda}$ [days]    & \multicolumn{3}{c}{$7.01^{+1.37}_{-1.31}$}  \\
ln Coherence parameter, $\ln{\Gamma}$  & \multicolumn{3}{c}{$1.8^{+2.4}_{-5.4}$} \\
ln Periodic timescale, $\ln{P_{\rm{GP}}}$ [days] & \multicolumn{3}{c}{$4.81^{1.79}_{-1.61}$} \\
Additive jitter, $s$ [m s$^{-1}$] & \multicolumn{3}{c}{$0.19^{+0.63}_{-0.04}$} \smallskip \\

& \emph{GJ 1132b} & \emph{GJ 1132c} & \emph{GJ 1132(d)} \\
\emph{Derived Parameters} & & & \smallskip \\
Period, $P$ [days] & $1.628931 \pm 0.000027$  & 8.929 $\pm$ 0.010 & 176.9 $\pm$ 5.1    \\
Time of inferior conjunction, $T_0$ [BJD-2,450,000]      & 7184.55786 $\pm$ 0.00031    & 7506.02 $\pm$ 0.34 & $7496.8^{+14.4}_{-8.6}$  \\
Radial velocity semi-amplitude, $K$ [m s$^{-1}$] & 2.85 $\pm$ 0.34   & 2.57 $\pm$ 0.39 & $3.03^{+0.58}_{-0.88}$    \\
$h=\sqrt{e}\cos{\omega}$ & $0.05 \pm 0.13$ & $-0.12^{+0.28}_{-0.25}$ & $-0.10^{+0.27}_{-0.28}$ \\
$k=\sqrt{e}\sin{\omega}$ & $-0.12 \pm 0.25$ & $0.14 \pm 0.24$ & $-0.05 \pm 0.29$ \\

\emph{Calculated Parameters} & & \smallskip \\
Semimajor axis, $a$ [AU]  & 0.0153 $\pm$ 0.0005 & 0.0476 $\pm$ 0.0017 & $0.35 \pm 0.01$  \\
Eccentricity, $e^{\ast}$    &  $< 0.22$ & $< 0.27$ & $<0.53$ \\
Planet mass, $M_{p}$ [$M_{\oplus}$] & 1.66 $\pm$ 0.23     & -  &  -      \\
Minimum planet mass, $M_{p}\sin{i}$ [$M_{\oplus}$]   & 1.66 $\pm$ 0.23     &  $2.64 \pm 0.44$ & $8.4^{+1.7}_{-2.5}$    \\
Planet density, $\rho_p$ [$\mathrm{g\;cm^{-3}}$]$^{\bullet}$ & 6.3 $\pm$ 1.3    & - & - \\
Surface gravity, $g$ [$\mathrm{m\;s^{-2}}$]$^{\bullet}$ & 12.9 $\pm$ 2.2 & - & -\\
Escape velocity, $v_{\rm{esc}}$[$\mathrm{km\;s^{-1}}$]$^{\bullet}$ &  13.6 $\pm$ 1.0  & - & - \\
Equilibrium temperature, $T_{\rm{eq}}$ [K] & & & \\
\hspace{2pt} Bond albedo of 0.3 (Earth-like) & 529 $\pm$ 9 & 300 $\pm$ 5 & 111 $\pm$ 2 \\
\hspace{2pt} Bond albedo of 0.75 (Venus-like) & 409 $\pm$ 7 & 232 $\pm$ 4 &  86 $\pm$ 1 \\

\hline
\end{tabular}
\begin{list}{}{}
\item {\bf{Notes.}} $^{(\bullet)}$ assuming a planetary radius of $1.13 \pm 0.056 R_{\oplus}$ \citep{2015Natur.527..204B}.
$^{(\ast)}$ upper limit, $95^{\rm th}$ percentile of the posterior PDF. 
$\Msun$~=~1.98842\ten[30]~kg, \Rsun~=~6.95508\ten[8]~m, 
\Mearth~=~5.9736\ten[24]~kg, \Rearth~=~6,378,137~m.
\end{list}
\end{table*}

\section{Discussions}

\subsection{GJ1132b in context}

Compared to the discovery paper, we revise the RV semi-amplitude
from $2.76 \pm 0.92$ to $2.85 \pm 0.34$ {\rm $m\,s^{-1}$}. This is faster a gain in precision than that expected from the larger number of RVs ($0.92/0.48\simeq\sqrt{128/25}$). We surmise this is because a 3-planet model is a more adequate description of the data.  This improves 
the mass determination by $\sim 45$\% from $1.62 \pm 0.55$ M$_{\oplus}$ to
$1.66 \pm 0.23$ M$_{\oplus}$. Together with the radius of GJ1132b measured by \citet{2017AJ....154..142D}, its bulk density then
becomes $5.9 \pm 1.9$ g\,cm$^{-3}$ and thus confirms its rocky nature. 

\begin{figure*}
\sidecaption
\includegraphics[width=12cm]{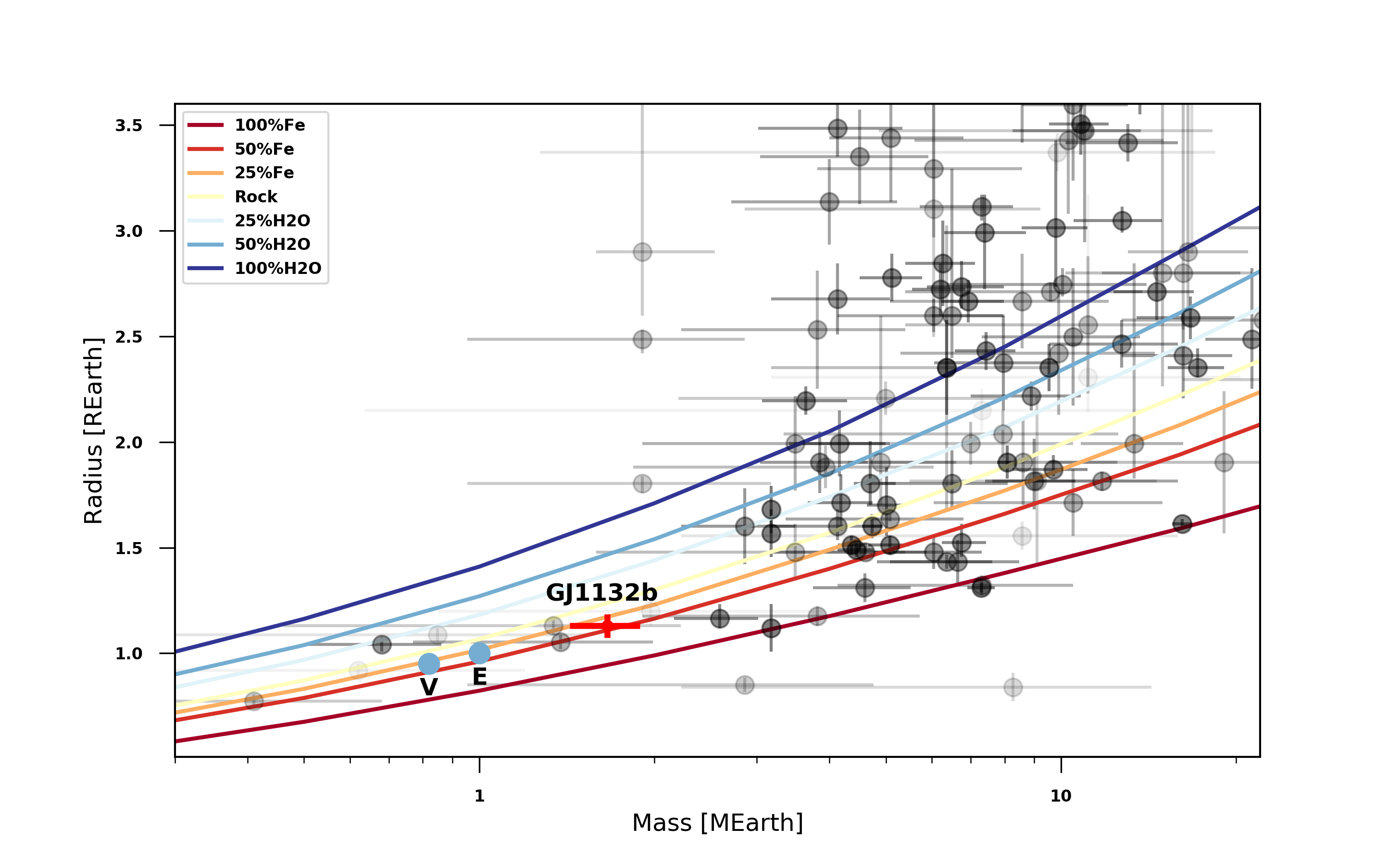}
\caption{Mass-radius diagram for planets with masses $M_p<20~M_\oplus$ and radii $R_p<3.5~R_\oplus$ queried from NASA Exoplanet Archive. Gray level scales linearly with relative uncertainties (with equal weight for both mass and radius). GJ1132b is shown in red whereas blue circles show Earth and Venus. Curves are for mass-radius relations predicted by theoretical models \citep{2013PASP..125..227Z}.}
       \label{fig:mr}
\end{figure*}

The mass-period diagram (Fig.~\ref{fig:mr}) places GJ1132b in context and compares its mass and radius both with other transit detection and with theoretical curves for different bulk compositions  \citep{2013PASP..125..227Z}. The density of GJ1132b appears compatible with a rocky or a denser compositions. With such a diagram, \citep{2015ApJ...801...41R} already observed that below $\sim$1.6~R$_\oplus$ planets are predominantly rocky, i.e. are preferentially found below the mass-radius curves that include significant water or lighter elements in their composition. And indeed, the largest planet that is more than 1-$\sigma$ away from the rocky curve is Kepler-60b, with a radius of 1.7~$R_\oplus$ \citep{Steffen:2013ja, JontofHutter:2016ch}.

Conversely, a similar threshold can now be observed with mass: to the exception of 3 planets with very large uncertainties \citep[Kepler -11b, -11f and -177b][]{2011Natur.470...53L,2013ApJ...770..131L, 2014ApJS..210...25X}, no planet is seen off the rocky curve by more than 1-$\sigma$ below a mass threshold of $\sim$3$ M_\oplus$.

\subsection{GJ1132c, a temperate super-Earth}

With a host star's luminosity $L_\star=0.00438\pm0.00034~L_\odot$  \citep{2017Natur.544..333D} and a semi-major $a = 0.048$ AU, GJ1132c receives about 1.9 times as much flux as Earth from our Sun. Its equilibrium temperature ranges from 232 K for a Bond albedo equals to that of Venus (A=0.75) and up to 328 K for a Bond albedo A=0. 

The most recent works that delineate the habitable zone around M dwarfs \citep[e.g.][]{2016ApJ...819...84K} place the inner edge for GJ1132 around 1.6 times the stellar radiation received by Earth. GJ1132c would thus be considered significantly too irradiated to have liquid water on its surface. Yet the planet remains of considerable interest in the context of habitability. The concept remains poorly understood and will remain so until inhabited worlds are actually found. The habitable-zone's inner edge is thus subject to change with future works. Also, providing future instrumentation were to reach sufficient contrast to resolve the planet from its parent star, probing the existence of an atmosphere would tell us how resilient this atmosphere can be against stellar irradiation, and thus more generally constrain the habitability of M-dwarf planets regardless of the habitability of GJ1132c itself. At a distance of 12 parsecs however \citep[][]{2015Natur.527..204B}, transmission and occultation spectroscopy are probably the sole methods able to resolve such an atmosphere, meaning GJ1132c would be required to transit.

\subsection{Transit search for planet c}

GJ1132c orbits at a distance of about 49$\pm$3 $R_\star$ from its host star and, without prior knowledge on the system orientation, the probability to see the planet in transit at inferior conjunction would be $\sim$1/50. We do nevertheless have prior knowledge on the system orientation since GJ1132 is already known to host a transiting planet with a measured orbital inclination of 88.68$^{+0.40}_{-0.33}$ degrees \citep{2017AJ....154..142D}. Considering only that nominal value (88.68 degrees), additional planets with strictly co-planar orbits would be seen to transit up to separations of $\sim43~R_\star$, and would be missed beyond. This limit is most probably inside the orbit of GJ1132c and, at first, one may think the prior knowledge we have from GJ1132b may actually nullify the probability of observing any transit for GJ1132c. This is, however, neglecting both the uncertainty on GJ1132c's orbital inclination and possible deviations from perfect coplanarity. 

We can instead include both uncertainty and non-coplanarity in our prior. Using the formalism of \citet{2010ApJ...712.1433B}, we make distributions of inclinations centered around 88.68 degrees and with various standard deviations. In Fig.~\ref{fig4}, one can see the prior probability that GJ1132c undergoes transit quickly jumps to $\sim$43\% for a distribution of inclinations with only a small standard deviation of 1 degree.

\begin{figure}
\includegraphics[width=0.9\linewidth]{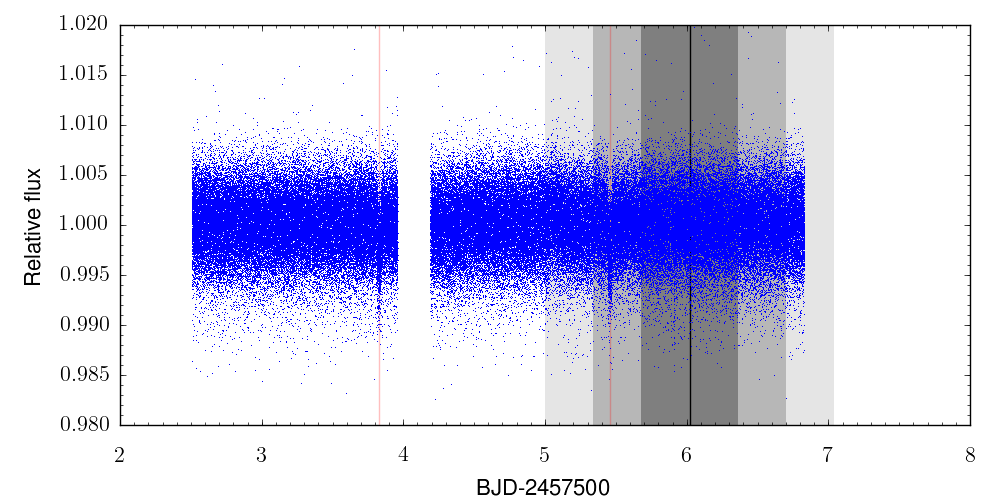}\\
\caption{\label{spitzer} Ephemerids for GJ1132c's inferior conjunction on top of Spitzer photometry. From this data set, \citet{2017AJ....154..142D} exclude possible transits of any additionnal planet larger than Mars. GJ1132 c transits are therefore largely excluded.}
       
\end{figure}

\begin{figure}
\includegraphics[width=0.9\linewidth]{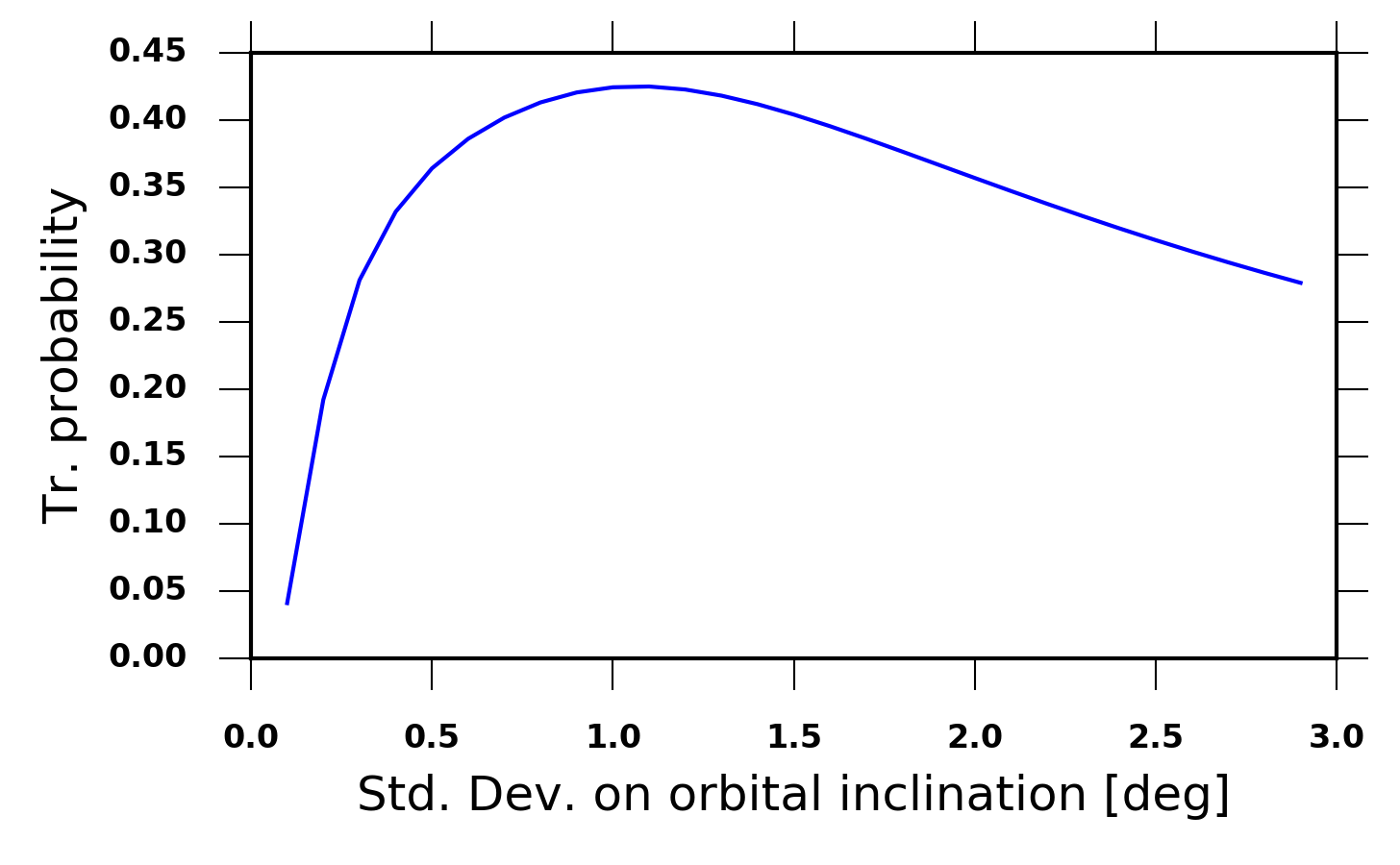}\\
\caption{\label{fig4}Transit probability of GJ1132c with gaussian prior on inclination. We chose the inclination of planet b for the central value of the distribution ($i_b = 88.68$ degrees) and we compute the transit probability for various standard deviations of this distribution (x axis) to reflect both the uncertainty on $i_b$ and small possible non-coplanarity.}
       
\end{figure}

\begin{figure*}
\includegraphics[width=0.9\linewidth]{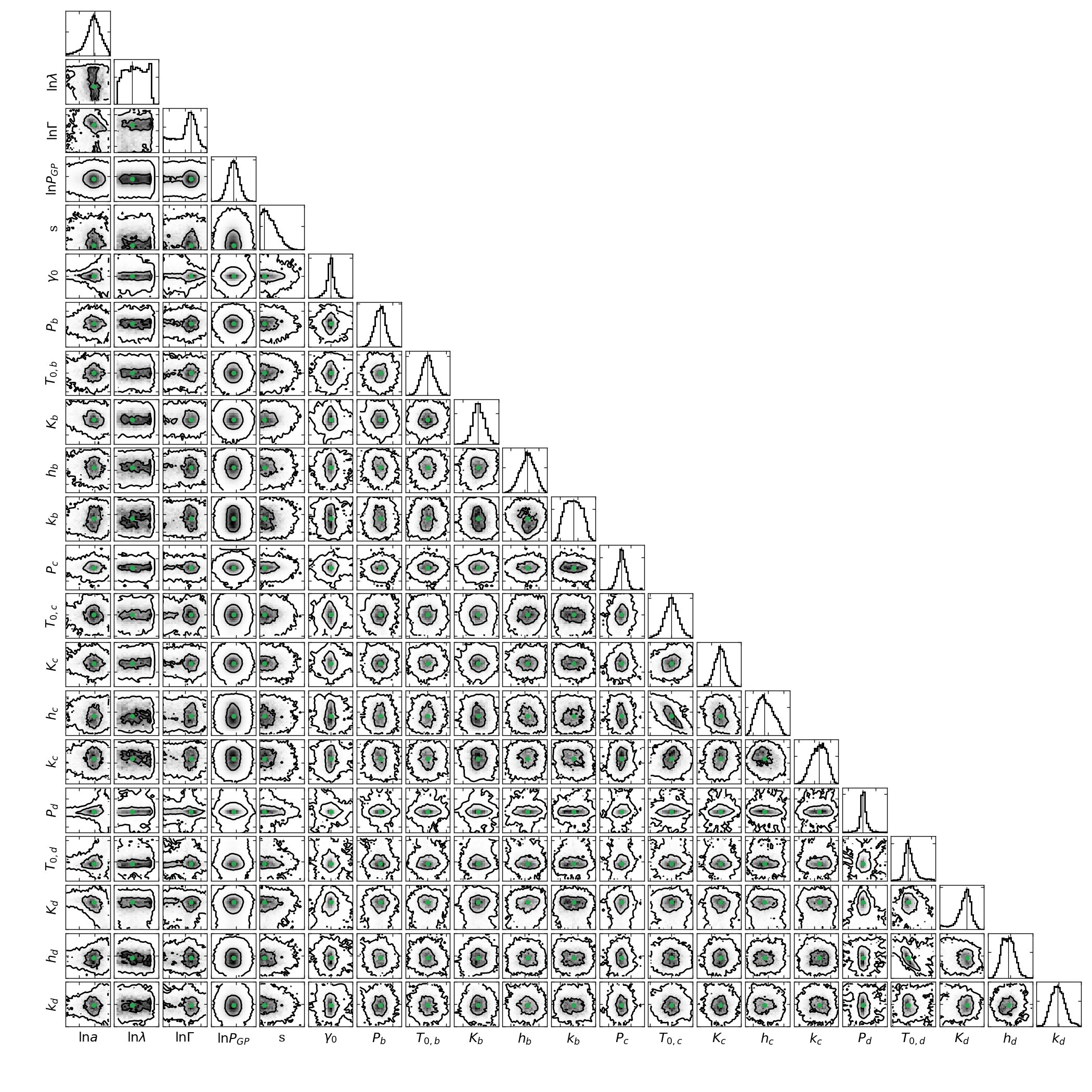}\\
\caption{Marginal posterior PDFs of the model parameters. See text for details.}
       \label{A1}
\end{figure*}

\begin{figure*}
\includegraphics[width=0.3\linewidth]{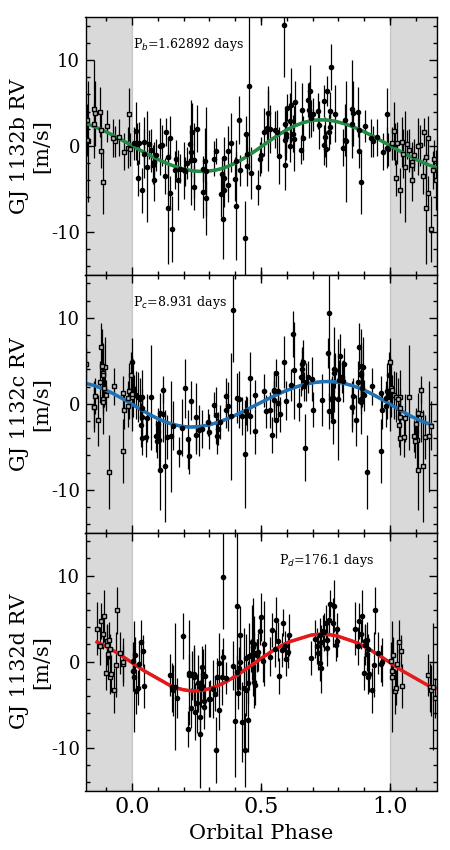}\\
\caption{Phase folded RV decomposition for the 3-planet model with GP regression removed.}
       \label{A2}
\end{figure*}

\begin{figure*}
\includegraphics[width=0.3\linewidth]{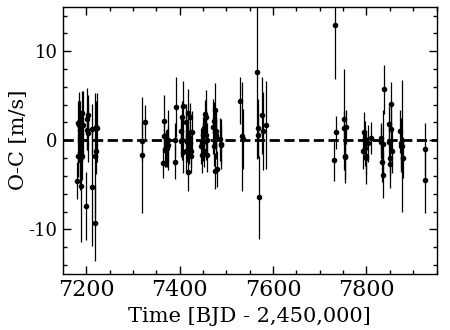}\\
\caption{Radial velocities residuals (O$-$C) as a function of time. }
       \label{A3}
\end{figure*}

From our analysis in Sect.~\ref{sect:analysis}, we derived an ephemeris for the passage of GJ1132c at inferior conjunction (BJD=$2457506.02\pm0.34$). 
As seen in Fig.~\ref{spitzer}, this falls inside the long 100-h monitoring made by \citet{2017AJ....154..142D} with {\it Spitzer}, which covers epochs between BJD=2457502.5 and 2457506.8 with almost no interruption and a sensitivity down to planets smaller than Mars \citep{2017AJ....154..142D}. Possible transits of GJ1132c are largely ruled out with $<1\%$ chance an existing transit was missed. 

The prior probability joined to the probability left by the incomplete coverage of the transit time-window gives a posterior probability $\le 0.43$\% that GJ1132c undergoes transits\footnote{0.43\% is the upper limit considering a standard deviation of 1 degree in our above calculations and for all other standard deviations the posterior probability is less than $0.43\%$.}. GJ1132c transits are almost fully ruled out.

\subsection{Stellar activity or a cold planet}

Based on activity diagnostics, we were not able to rule out stellar activity as the main cause of the 170-d radial velocity periodicity. However, if later observations were to confirm the planetary nature of that signal, GJ1132 d would have a minimum mass $m_d \sin i = 8.4_{-2.5}^{+1.7}$~\Mearth, in a mass domain between super-Earths and mini-Neptunes. Its semi-major axis $a_d = 0.35\pm0.01$ AU would place this planet beyond the ice line, with an equilibrium temperature of 86 K (resp 111 K) for a Bond albedo of 0.75 (resp. 0.3). Planet $d$ would be $\sim$7.3 times further away from the star than planet $c$. A priori, its transit probability would thus be $\sim7.3$ smaller. The $Spitzer$ photometry rejects only a small fraction (13\%) of possible transit configurations.

\subsection{Transit-timing variations}

The orbital periods ratio of planets b and c is close to 11/2. Even if inside a resonance, its low order would imply low transit-timing variations. We use the {\sc \tt Rebound} code \citep{2011ascl.soft10016R} with the {\sc \tt WHFast} integrator \citep{2015MNRAS.452..376R} to compute TTVs for this system. Although not shown here, TTVs were found to be generally less than 30s, in agreement with the lack of TTVs found in \citet{2017AJ....154..142D}.\\

\section{Conclusion}

To conclude, our HARPS radial-velocity follow-up helps to bring a more complete, more complex description on the GJ1132 system. We confirmed the detection of planet 'b' with the sole radial velocities, we refined its characteristics including its mass, density and eccentricity. We also detect at least one new planet in the system. Assuming coplanarity with planet 'b', GJ1132 c is a super-Earth with mass $m_c = 2.75^{+0.76}_{-0.61}$~M$_\oplus$. Its equilibrium temperature is that of a temperate planet although it is probably too close to the star to allow for liquid water on its surface. Finally, we also detect a third keplerian signal but leave its true nature --planet or stellar activity-- undecided. If confirmed as a planet, GJ1132 d would be a super-Earth or mini-Neptune found further away from the star and beyond the ice line. Further observations, and potentially at other wavelengths with infrared spectrographs such as \textsc{Carmenes} \citep{2014SPIE.9147E..1FQ, 2018arXiv180500830S} or \textsc{SPIRou} \citep{2013sf2a.conf..497D}, may help further interpret the true nature of that signal.

\begin{acknowledgements} We are grateful to our referee for comments that significantly improved our manuscript.
XB, JMA and AW acknowledge funding from the European Research Council under the ERC Grant Agreement n. 337591-ExTrA. RC and RD acknowlege financial support from the National Research Council of Canada and the Institute for Research on Exoplanets. NCS acknowledges the support by Funda\c{c}\~ao para a Ci\^encia e a Tecnologia (FCT, Portugal) through the research grant through national funds and by FEDER through COMPETE2020 by grants UID/FIS/04434/2013\&POCI-01-0145-FEDER-007672 and PTDC\-/FIS-AST/1526/2014\&POCI-01-0145-FEDER-016886, as well as through Investigador FCT contract nr. IF/00169/2012/CP0150/CT0002. JAD gratefully acknowledges funding from the Heising-Simons Foundation's 51 Pegasi b postdoctoral fellowship. The MEarth Team gratefully acknowledges funding from the David and Lucille Packard Fellowship for Science and Engineering (awarded to D.C.). This material is based upon work supported by the National Science Foundation under grants AST-0807690, AST-1109468, AST-1004488 (Alan T. Waterman Award), and AST-1616624. E.R.N. is supported by an NSF Astronomy and Astrophysics Postdoctoral Fellowship. This publication was made possible through the support of a grant from the John Templeton Foundation. The opinions expressed in this publication are those of the authors and do not necessarily reflect the views of the John Templeton Foundation. This research has made use of the NASA Exoplanet Archive, which is operated by the California Institute of Technology, under contract with the National Aeronautics and Space Administration under the Exoplanet Exploration Program.
\end{acknowledgements}

\bibliographystyle{aa}
\bibliography{mybib}

\onecolumn

\longtab{3}{\begin{longtable}{lllllllllll}
\caption{\label{tab:rv} HARPS RV and spectroscopic indices time-series. Radial-velocities are given in the solar system barycentric reference frame. }\\

\hline\hline
BJD-2400000.0       & RV [km/s]       & $\sigma$RV [km/s] &H$_\alpha$     &$\sigma$H$_\alpha$&H$_\beta$     &$\sigma$H$_\beta$&NaD&$\sigma$NaD& S & $\sigma$S\\\hline \endfirsthead
\caption{continued.}\\
\hline\hline
BJD-2400000.0       & RV [km/s]       & $\sigma$RV [km/s] &H$_\alpha$     &$\sigma$H$_\alpha$&H$_\beta$     &$\sigma$H$_\beta$&NaD&$\sigma$NaD& S & $\sigma$S\\\hline \endhead \hline \endfoot

57180.4657      &35.0715        &0.0021 &0.0736 &0.0002 &0.0594 &0.0006 &0.0078 &0.0002 &1.1837 &0.1395\\
57181.4875      &35.0777        &0.0024 &0.0737 &0.0003 &0.0593 &0.0008 &0.0073 &0.0003 &1.2959 &0.1978\\
57182.5211      &35.0790        &0.0024 &0.0738 &0.0003 &0.0604 &0.0008 &0.0072 &0.0003 &1.1192 &0.2717\\
57183.5369      &35.0766        &0.0029 &0.0733 &0.0003 &0.0628 &0.0010 &0.0075 &0.0004 &1.0605 &0.2463\\
57184.5827      &35.0770        &0.0034 &0.0746 &0.0004 &0.0623 &0.0013 &0.0081 &0.0005 &1.3121 &0.4228\\
57185.4828      &35.0725        &0.0034 &0.0774 &0.0004 &0.0697 &0.0012 &0.0083 &0.0005 &1.1838 &0.2943\\
57186.4758      &35.0688        &0.0034 &0.0789 &0.0004 &0.0730 &0.0012 &0.0090 &0.0005 &1.0703 &0.3002\\
57187.4707      &35.0770        &0.0033 &0.0739 &0.0004 &0.0611 &0.0011 &0.0076 &0.0005 &1.2266 &0.2843\\
57188.4751      &35.0677        &0.0063 &0.0759 &0.0007 &0.0639 &0.0023 &0.0077 &0.0011 &0.5631 &0.5589\\
57189.4766      &35.0767        &0.0024 &0.0737 &0.0003 &0.0620 &0.0009 &0.0079 &0.0003 &1.3587 &0.3760\\
57190.4747      &35.0828        &0.0025 &0.0757 &0.0003 &0.0666 &0.0009 &0.0087 &0.0003 &1.1962 &0.2160\\
57191.4635      &35.0733        &0.0026 &0.0726 &0.0003 &0.0579 &0.0008 &0.0072 &0.0003 &1.0801 &0.2145\\
57192.4695      &35.0811        &0.0038 &0.0756 &0.0004 &0.0667 &0.0012 &0.0093 &0.0006 &1.2518 &0.3017\\
57199.4737      &35.0686        &0.0039 &0.0759 &0.0003 &0.0692 &0.0008 &0.0149 &0.0003 &0.8405 &0.2049\\
57200.4709      &35.0826        &0.0022 &0.0730 &0.0003 &0.0646 &0.0007 &0.0076 &0.0002 &1.1800 &0.1595\\
57201.4700      &35.0781        &0.0035 &0.0759 &0.0004 &0.0736 &0.0013 &0.0108 &0.0004 &0.8145 &0.3528\\
57203.4724      &35.0772        &0.0033 &0.0767 &0.0004 &0.0679 &0.0010 &0.0097 &0.0004 &2.0155 &0.2546\\
57204.4690      &35.0739        &0.0022 &0.0732 &0.0003 &0.0602 &0.0007 &0.0088 &0.0002 &0.6547 &0.1638\\
57211.4898      &35.0786        &0.0028 &0.1089 &0.0004 &0.1535 &0.0014 &0.0181 &0.0004 &3.7218 &0.3768\\
57212.4771      &35.0678        &0.0065 &0.0759 &0.0007 &0.0694 &0.0025 &0.0085 &0.0010 &3.5179 &4.3547\\
57218.4902      &35.0808        &0.0070 &0.0774 &0.0007 &0.0719 &0.0027 &0.0072 &0.0012 &2.6520 &1.2886\\
57219.4812      &35.0686        &0.0043 &0.0766 &0.0005 &0.0763 &0.0016 &0.0071 &0.0006 &1.3183 &0.6870\\
57220.4694      &35.0780        &0.0029 &0.0751 &0.0003 &0.0669 &0.0010 &0.0074 &0.0003 &0.9122 &0.3073\\
57221.4706      &35.0769        &0.0025 &0.0769 &0.0003 &0.0660 &0.0009 &0.0079 &0.0003 &1.3581 &0.3381\\
57222.4842      &35.0738        &0.0039 &0.0784 &0.0004 &0.0762 &0.0016 &0.0109 &0.0005 &2.8997 &0.6544\\
57318.8586      &35.0770        &0.0022 &0.0778 &0.0003 &0.0709 &0.0007 &0.0101 &0.0002 &1.4137 &0.1460\\
57319.8553      &35.0739        &0.0065 &0.0747 &0.0007 &0.0550 &0.0022 &0.0090 &0.0010 &0.9849 &0.9136\\
57325.8585      &35.0855        &0.0019 &0.0747 &0.0002 &0.0637 &0.0006 &0.0084 &0.0002 &1.3391 &0.0847\\
57364.8062      &35.0725        &0.0017 &0.0732 &0.0002 &0.0567 &0.0005 &0.0053 &0.0001 &0.8243 &0.0849\\
57365.8194      &35.0703        &0.0016 &0.0736 &0.0002 &0.0563 &0.0004 &0.0057 &0.0001 &0.7972 &0.0720\\
57366.7942      &35.0782        &0.0029 &0.0739 &0.0003 &0.0582 &0.0008 &0.0057 &0.0003 &0.8061 &0.1615\\
57367.8072      &35.0745        &0.0020 &0.0734 &0.0002 &0.0551 &0.0005 &0.0057 &0.0002 &0.8998 &0.1020\\
57370.8365      &35.0741        &0.0021 &0.0721 &0.0002 &0.0564 &0.0006 &0.0065 &0.0002 &1.0348 &0.1061\\
57371.8364      &35.0755        &0.0022 &0.0736 &0.0003 &0.0577 &0.0006 &0.0060 &0.0002 &0.6316 &0.1142\\
57373.8390      &35.0704        &0.0034 &0.0740 &0.0004 &0.0571 &0.0009 &0.0059 &0.0004 &0.6013 &0.1954\\
57374.7626      &35.0756        &0.0020 &0.0746 &0.0002 &0.0601 &0.0006 &0.0072 &0.0002 &0.9275 &0.1048\\
57389.8172      &35.0778        &0.0019 &0.0740 &0.0002 &0.0606 &0.0005 &0.0064 &0.0002 &0.8959 &0.0829\\
57390.8011      &35.0757        &0.0019 &0.0740 &0.0002 &0.0616 &0.0005 &0.0065 &0.0002 &0.9732 &0.0809\\
57391.8098      &35.0764        &0.0034 &0.0741 &0.0004 &0.0585 &0.0010 &0.0071 &0.0004 &0.3834 &0.2935\\
57401.7321      &35.0757        &0.0015 &0.0752 &0.0002 &0.0649 &0.0004 &0.0075 &0.0001 &1.1823 &0.0609\\
57403.7974      &35.0823        &0.0019 &0.0764 &0.0002 &0.0685 &0.0005 &0.0083 &0.0002 &1.3221 &0.0755\\
57404.7412      &35.0796        &0.0022 &0.0790 &0.0003 &0.0730 &0.0007 &0.0083 &0.0002 &1.4257 &0.1107\\
57405.6930      &35.0881        &0.0017 &0.0751 &0.0002 &0.0677 &0.0005 &0.0082 &0.0002 &1.2337 &0.0754\\
57406.7642      &35.0841        &0.0028 &0.0754 &0.0003 &0.0681 &0.0009 &0.0077 &0.0003 &1.4073 &0.1748\\
57407.6980      &35.0793        &0.0024 &0.0737 &0.0002 &0.0505 &0.0045 &0.0080 &0.0003 &-1.8703        &3.2426\\
57412.7474      &35.0790        &0.0023 &0.0757 &0.0003 &0.0635 &0.0007 &0.0085 &0.0002 &0.9553 &0.1079\\
57413.7202      &35.0876        &0.0019 &0.0733 &0.0002 &0.0627 &0.0006 &0.0086 &0.0002 &1.1364 &0.0899\\
57415.7809      &35.0830        &0.0022 &0.0772 &0.0003 &0.0704 &0.0007 &0.0089 &0.0002 &1.6911 &0.1186\\
57416.7217      &35.0842        &0.0027 &0.0737 &0.0003 &0.0656 &0.0008 &0.0076 &0.0003 &1.3785 &0.1412\\
57417.7195      &35.0773        &0.0023 &0.0763 &0.0003 &0.0693 &0.0007 &0.0083 &0.0002 &1.6400 &0.1145\\
57418.7193      &35.0775        &0.0021 &0.0742 &0.0002 &0.0663 &0.0006 &0.0084 &0.0002 &1.3759 &0.1162\\
57420.7413      &35.0797        &0.0019 &0.0929 &0.0003 &0.1135 &0.0007 &0.0137 &0.0002 &2.6172 &0.1104\\
57421.7295      &35.0863        &0.0016 &0.0740 &0.0002 &0.0647 &0.0005 &0.0078 &0.0001 &1.1015 &0.0828\\
57422.7158      &35.0802        &0.0017 &0.0797 &0.0002 &0.0772 &0.0005 &0.0102 &0.0002 &1.8171 &0.0822\\
57423.7243      &35.0843        &0.0019 &0.0778 &0.0002 &0.0739 &0.0006 &0.0096 &0.0002 &1.7806 &0.1194\\
57424.6908      &35.0800        &0.0018 &0.0734 &0.0002 &0.0627 &0.0005 &0.0080 &0.0002 &1.2936 &0.0928\\
57425.6801      &35.0815        &0.0024 &0.0744 &0.0003 &0.0640 &0.0007 &0.0080 &0.0003 &1.6742 &0.1567\\
57446.6185      &35.0818        &0.0020 &0.0798 &0.0002 &0.0753 &0.0006 &0.0102 &0.0002 &1.4571 &0.0769\\
57447.6264      &35.0804        &0.0021 &0.0772 &0.0003 &0.0716 &0.0006 &0.0098 &0.0002 &1.6056 &0.0810\\
57448.5920      &35.0821        &0.0019 &0.0764 &0.0002 &0.0691 &0.0006 &0.0085 &0.0002 &1.5313 &0.0688\\
57449.5756      &35.0875        &0.0022 &0.1053 &0.0003 &0.1542 &0.0010 &0.0171 &0.0003 &2.6527 &0.0831\\
57450.5911      &35.0831        &0.0021 &0.0754 &0.0002 &0.0693 &0.0006 &0.0084 &0.0002 &1.1410 &0.0705\\
57451.5918      &35.0841        &0.0018 &0.0733 &0.0002 &0.0640 &0.0005 &0.0079 &0.0002 &1.1318 &0.0719\\
57452.5529      &35.0839        &0.0018 &0.0819 &0.0002 &0.0872 &0.0006 &0.0099 &0.0002 &1.8971 &0.0765\\
57453.5449      &35.0816        &0.0016 &0.0742 &0.0002 &0.0654 &0.0005 &0.0084 &0.0001 &1.3474 &0.0716\\
57455.7948      &35.0818        &0.0028 &0.0759 &0.0003 &0.0679 &0.0010 &0.0086 &0.0003 &1.2791 &0.3333\\
57456.5705      &35.0851        &0.0030 &0.0760 &0.0003 &0.0674 &0.0010 &0.0091 &0.0004 &1.4816 &0.2336\\
57457.5540      &35.0837        &0.0019 &0.0742 &0.0002 &0.0654 &0.0006 &0.0082 &0.0002 &1.4374 &0.0909\\
57458.5628      &35.0821        &0.0019 &0.0735 &0.0002 &0.0658 &0.0006 &0.0081 &0.0002 &1.2384 &0.0941\\
57470.6955      &35.0846        &0.0031 &0.0721 &0.0004 &0.0588 &0.0009 &0.0093 &0.0004 &0.7902 &0.1269\\
57472.7555      &35.0793        &0.0022 &0.0751 &0.0003 &0.0670 &0.0007 &0.0081 &0.0002 &0.8746 &0.1102\\
57473.5981      &35.0805        &0.0021 &0.0740 &0.0002 &0.0638 &0.0006 &0.0081 &0.0002 &1.0454 &0.1138\\
57474.6150      &35.0808        &0.0017 &0.0724 &0.0002 &0.0605 &0.0005 &0.0069 &0.0002 &0.9109 &0.0870\\
57475.6271      &35.0885        &0.0030 &0.0723 &0.0004 &0.0620 &0.0008 &0.0072 &0.0004 &0.9834 &0.1501\\
57476.6011      &35.0770        &0.0020 &0.0753 &0.0002 &0.0642 &0.0006 &0.0080 &0.0002 &1.3693 &0.1018\\
57477.5256      &35.0855        &0.0024 &0.0762 &0.0003 &0.0652 &0.0007 &0.0082 &0.0003 &2.1576 &0.1534\\
57477.6036      &35.0853        &0.0019 &0.0767 &0.0002 &0.0679 &0.0006 &0.0087 &0.0002 &1.4413 &0.1070\\
57478.6655      &35.0835        &0.0022 &0.0724 &0.0002 &0.0600 &0.0006 &0.0065 &0.0002 &1.2298 &0.1417\\
57479.6303      &35.0742        &0.0020 &0.0738 &0.0002 &0.0619 &0.0006 &0.0068 &0.0002 &1.7087 &0.1287\\
57486.5663      &35.0803        &0.0024 &0.0767 &0.0003 &0.0711 &0.0008 &0.0080 &0.0003 &1.3607 &0.1315\\
57488.5522      &35.0801        &0.0021 &0.0726 &0.0003 &0.0602 &0.0006 &0.0070 &0.0002 &0.9245 &0.0871\\
57488.7183      &35.0808        &0.0028 &0.0731 &0.0003 &0.0622 &0.0009 &0.0070 &0.0003 &1.3646 &0.1755\\
57529.5196      &35.0866        &0.0027 &0.0739 &0.0003 &0.0632 &0.0008 &0.0083 &0.0003 &0.9024 &0.1630\\
57533.5194      &35.0734        &0.0061 &0.0816 &0.0007 &0.0760 &0.0022 &0.0114 &0.0011 &1.1707 &0.4517\\
57536.5792      &35.0743        &0.0033 &0.0802 &0.0004 &0.0881 &0.0013 &0.0097 &0.0005 &1.4100 &0.2996\\
57566.4667      &35.0868        &0.0098 &0.0723 &0.0010 &0.0665 &0.0041 &0.0090 &0.0019 &1.2531 &0.6473\\
57567.4772      &35.0788        &0.0025 &0.0758 &0.0003 &0.0740 &0.0009 &0.0094 &0.0003 &1.5649 &0.2308\\
57568.4726      &35.0819        &0.0033 &0.0747 &0.0003 &0.0719 &0.0013 &0.0095 &0.0005 &1.2613 &0.3376\\
57569.5552      &35.0673        &0.0047 &0.0743 &0.0005 &0.0709 &0.0019 &0.0085 &0.0008 &1.5105 &0.4358\\
57576.5030      &35.0850        &0.0044 &0.0803 &0.0005 &0.0746 &0.0016 &0.0092 &0.0007 &1.3251 &0.2503\\
57577.5167      &35.0769        &0.0044 &0.0735 &0.0004 &0.0637 &0.0017 &0.0087 &0.0007 &1.4878 &0.6083\\
57584.5054      &35.0840        &0.0049 &0.0731 &0.0005 &0.0687 &0.0019 &0.0088 &0.0008 &0.6917 &0.3945\\
57729.8316      &35.0752        &0.0023 &0.0730 &0.0003 &0.0620 &0.0007 &0.0076 &0.0002 &1.1951 &0.1406\\
57732.8377      &35.0890        &0.0060 &0.0779 &0.0007 &0.0654 &0.0020 &0.0085 &0.0009 &1.5138 &0.9210\\
57734.8518      &35.0813        &0.0019 &0.0734 &0.0002 &0.0628 &0.0005 &0.0075 &0.0002 &1.2762 &0.0887\\
57751.7936      &35.0771        &0.0019 &0.0724 &0.0002 &0.0620 &0.0006 &0.0081 &0.0002 &1.2669 &0.1141\\
57752.8122      &35.0847        &0.0057 &0.0729 &0.0006 &0.0644 &0.0019 &0.0075 &0.0009 &1.1037 &0.4191\\
57753.8014      &35.0781        &0.0030 &0.0724 &0.0003 &0.0642 &0.0009 &0.0084 &0.0004 &1.2188 &0.2001\\
57754.8207      &35.0770        &0.0024 &0.0723 &0.0003 &0.0621 &0.0007 &0.0078 &0.0003 &1.1402 &0.1332\\
57755.8231      &35.0831        &0.0019 &0.0751 &0.0002 &0.0682 &0.0006 &0.0091 &0.0002 &1.3481 &0.1158\\
57792.7277      &35.0751        &0.0020 &0.0764 &0.0002 &0.0728 &0.0006 &0.0092 &0.0002 &1.3278 &0.1112\\
57795.6726      &35.0782        &0.0022 &0.0743 &0.0003 &0.0647 &0.0006 &0.0096 &0.0002 &1.0836 &0.0807\\
57796.6455      &35.0843        &0.0021 &0.0829 &0.0003 &0.0810 &0.0007 &0.0110 &0.0002 &1.6221 &0.0927\\
57797.6402      &35.0795        &0.0019 &0.0716 &0.0002 &0.0591 &0.0005 &0.0078 &0.0002 &1.0547 &0.0885\\
57798.6965      &35.0809        &0.0021 &0.0727 &0.0002 &0.0605 &0.0006 &0.0085 &0.0002 &1.1639 &0.0922\\
57799.6246      &35.0818        &0.0030 &0.0719 &0.0003 &0.0581 &0.0008 &0.0088 &0.0004 &1.1241 &0.1988\\
57803.7585      &35.0763        &0.0021 &0.0729 &0.0002 &0.0614 &0.0006 &0.0082 &0.0002 &1.1774 &0.1587\\
57810.6343      &35.0769        &0.0018 &0.0746 &0.0002 &0.0709 &0.0005 &0.0082 &0.0002 &1.4949 &0.0846\\
57830.5833      &35.0794        &0.0019 &0.0803 &0.0002 &0.0830 &0.0006 &0.0099 &0.0002 &1.9467 &0.1141\\
57831.5423      &35.0770        &0.0017 &0.0736 &0.0002 &0.0662 &0.0005 &0.0080 &0.0002 &1.1967 &0.1104\\
57832.5735      &35.0811        &0.0022 &0.0761 &0.0003 &0.0698 &0.0007 &0.0084 &0.0002 &1.5973 &0.1742\\
57834.5604      &35.0775        &0.0026 &0.0751 &0.0003 &0.0694 &0.0008 &0.0093 &0.0003 &1.5597 &0.2012\\
57835.5218      &35.0825        &0.0038 &0.0836 &0.0004 &0.0830 &0.0014 &0.0115 &0.0006 &1.5755 &0.3235\\
57836.5391      &35.0818        &0.0028 &0.0743 &0.0003 &0.0664 &0.0009 &0.0079 &0.0004 &1.3216 &0.2138\\
57847.6649      &35.0737        &0.0023 &0.0749 &0.0003 &0.0661 &0.0007 &0.0084 &0.0002 &1.2955 &0.2598\\
57848.7030      &35.0808        &0.0021 &0.0758 &0.0002 &0.0659 &0.0007 &0.0086 &0.0002 &2.0690 &0.3198\\
57849.7216      &35.0725        &0.0026 &0.0755 &0.0003 &0.0682 &0.0009 &0.0084 &0.0003 &1.1287 &0.4710\\
57850.5931      &35.0790        &0.0020 &0.0771 &0.0002 &0.0702 &0.0006 &0.0098 &0.0002 &1.4973 &0.1231\\
57851.7008      &35.0822        &0.0026 &0.0755 &0.0003 &0.0657 &0.0009 &0.0091 &0.0003 &1.1375 &0.2224\\
57852.6557      &35.0815        &0.0025 &0.0788 &0.0003 &0.0770 &0.0009 &0.0100 &0.0003 &1.0996 &0.2015\\
57854.6927      &35.0733        &0.0026 &0.0764 &0.0003 &0.0674 &0.0009 &0.0101 &0.0003 &1.0953 &0.1944\\
57872.5444      &35.0741        &0.0024 &0.0726 &0.0003 &0.0618 &0.0007 &0.0076 &0.0003 &1.1265 &0.1617\\
57873.6706      &35.0727        &0.0029 &0.0735 &0.0003 &0.0637 &0.0011 &0.0074 &0.0004 &0.9222 &0.3373\\
57874.5623      &35.0746        &0.0030 &0.0783 &0.0003 &0.0736 &0.0010 &0.0097 &0.0004 &1.4706 &0.2621\\
57875.6886      &35.0714        &0.0074 &0.0747 &0.0008 &0.0614 &0.0031 &0.0086 &0.0013 &1.3750 &0.7009\\
57877.5748      &35.0749        &0.0023 &0.0734 &0.0003 &0.0613 &0.0008 &0.0080 &0.0003 &1.1110 &0.2208\\
57924.5360      &35.0747        &0.0030 &0.0737 &0.0003 &0.0620 &0.0011 &0.0078 &0.0004 &2.0765 &0.3792\\
57925.5331      &35.0745        &0.0037 &0.0774 &0.0004 &0.0698 &0.0014 &0.0091 &0.0005 &1.8617 &0.4392\\
\end{longtable}}

\end{document}